# Specification-Driven Data Architecture Reconstruction: From Physical Code to Logical and Conceptual Specifications

Oleg Grynets
EPAM Systems
McLean, Virginia, USA
oleg_grynets@epam.com

Olena Pochernina
EPAM Systems,
Lviv, Ukraine
olena_pochernina@epam.com

Vasyl Lyashkevych
EPAM Systems
Lviv, Ukraine
vasyl_lyashkevych@epam.com

***Abstract*—Legacy database migrations often begin with incomplete or outdated documentation, leaving physical data definition language (DDL) as the principal evidence of data architecture. However, DDL does not fully encode conceptual intent, and model-generated completions can be plausible without being correct. This study proposes and evaluates a provenance-aware, deterministic-first pipeline for reconstructing logical and conceptual data specifications from Oracle-oriented DDL while explicitly separating observed facts, deterministic derivations, and large language model (LLM) suggestions. The pipeline performs DDL investigation, parsing, consolidation, primary-key backfilling, type normalization, and declared relationship-graph construction before optional LLM-assisted enrichment. It preserves source provenance in the deterministic catalog and declared relationship graph, and records inferred primary-key and foreign-key candidates in a separate reviewable overlay. We evaluated the implementation on 249 artifactized schema samples comprising 1,225 SQL files. The pipeline completed 244 samples (97.99%); 52 completed samples contained no extractable DDL. Across completed samples, the deterministic path reconstructed 208 tables and recovered 278 declared foreign-key records; 168 of the reconstructed tables lacked an explicitly parsed primary key before backfilling. LLM enrichment generated 100 foreign-key candidates in 36 samples, but the parent-table admissibility rate was only 17.9% for logical-specification candidates and 17.5% for conceptual-specification candidates. These findings show that the proposed deterministic-first architecture can preserve an auditable structural baseline, quantify observed primary-key and relationship gaps, and prevent model-generated hypotheses from being silently promoted to source-grounded architectural facts.**

***Keywords—large language models, database reverse engineering, Oracle DDL, schema reconstruction, specification-driven modernization, provenance, primary-key inference, foreign-key inference, schema migration.***

## I. Introduction

Modernization of enterprise software frequently depends on migration of legacy database schemas to new platforms, integration stacks, and security baselines. In such projects, the source DDL and related schema artifacts are often the most reliable technical evidence available when legacy documentation is missing, outdated, or misleading [1]–[3]. As a result, migration quality depends on how accurately one can recover the intended data architecture from physical schema statements rather than from complete architectural specifications [1]–[3].

This recovery process is difficult for two reasons. First, legacy schemas commonly under-specify architectural intent: not all conceptual constraints are necessarily translated into enforceable schema constraints, and important semantics can be "discarded" from the DBMS-managed schema and pushed into application logic, procedures, or other mechanisms outside the DBMS [1], [3]. Second, physical data representations are platform-specific, and direct translation to another platform is not one-to-one. More generally, direct transformation can preserve surface form while introducing semantic drift or hidden behavioral changes, which motivates reconstruction through an intermediate specification rather than purely syntactic conversion [4].

Recent specification-driven studies strengthen this interpretation by treating the specification as an operational intermediate representation rather than as passive documentation. A context-aware formal model describes such a representation through explicit goals, domain vocabulary, functional and data constraints, architectural information, validation rules, and traceability links [5], while evolution-aware specification-driven synthesis demonstrates how structured specifications can serve as regenerable control artifacts for downstream software generation [6]. More recent work further formalizes specification content as dependency-connected semantic blocks with explicit rules, decision points, and unresolved questions, emphasizing that underdetermined information should remain identifiable rather than being silently completed [7].

These challenges can be understood through the classic three-layer view of data architecture: physical, logical, and conceptual. The physical layer captures platform-bound storage constructs; the logical layer captures entities, attributes, keys, and relationships in a platform-independent form; and the conceptual layer captures domain-level abstractions and associations. Legacy DDL primarily describes the physical layer. Therefore, producing migration-ready documentation is a reverse-engineering problem: one must infer higher-level architecture from lower-level declarations while preserving traceability to source evidence [1]–[3].

This problem has long been recognized in database reverse engineering. Classical studies show that conceptual information may be distributed across schema declarations, naming conventions, data instances, and application-level logic, so reconstructing higher-level models requires both explicit evidence and carefully controlled elicitation of implicit structures [8]–[10]. Reverse-engineering approaches based on EER recovery and model transformations similarly show that moving from relational implementation artifacts to conceptual models is inherently a transformation across abstraction levels rather than a direct syntactic projection [11], [12]. These observations are particularly relevant when DDL is the principal available source because an absent physical constraint does not necessarily imply an absent domain relationship.

Prior research addresses important parts of this problem, including database reverse engineering (recovering conceptual schemas and constraints from relational artifacts) [1]–[3], [13], [14], and schema matching across heterogeneous schemas and systems [15]. However, a migration workflow still needs to distinguish source evidence from hypotheses and to expose the points at which parsing or inference fails. Without that distinction, a plausible reconstruction can be mistaken for a verified one, allowing unverifiable assumptions to propagate into downstream system design. The specific research gap is the absence of a provenance-aware reconstruction process that jointly preserves physical evidence, deterministic derivations, unresolved information, and model-generated hypotheses while transforming incomplete DDL into logical and conceptual specifications [4], [16].

This work addresses that gap with a staged framework for reconstructing logical and conceptual specifications from Oracle-oriented physical schemas. The pipeline performs DDL investigation, parsing, consolidation, deterministic backfilling, and graph construction before applying model-assisted enrichment. Its outputs preserve a declared graph separately from an enrichment overlay: declared and deterministically derived properties retain source information, while inferred primary-key and foreign-key candidates are recorded as suggestions with confidence and rationale. Missing structure remains representable rather than being silently converted into a declared fact. This design supports review and selective acceptance but does not by itself establish that every inferred element is semantically correct.

The central objective of this study is to determine whether incomplete Oracle-oriented DDL can be transformed into migration-oriented logical and conceptual specifications while preserving the epistemic status and available provenance metadata of reconstructed architectural elements. The evaluation therefore considers both execution behavior and the structure of produced outputs, including completion, empty-result, failure, key, relationship, and provenance patterns. Semantic correctness of inferred keys and relationships is treated as a separate validation question requiring known or independently reviewed reference schemas. The approach aims to make uncertainty visible, separate evidence from hypothesis, and provide a practical basis for controlled modernization decisions [4], [17].

To operationalize this objective, the study addresses the following research tasks:

- Define a representation that captures reconstructed logical and conceptual structure while preserving links to source DDL evidence.
- Design a multi-stage pipeline that combines parser outputs, rule-based recovery, and model-driven inference in a controlled workflow.
- Distinguish declared, deterministically derived, and model-suggested architectural elements using explicit provenance, source class, and confidence annotations.
- Evaluate pipeline completion, structural preservation, unresolved-information coverage, relationship reconstruction, and the availability of provenance metadata across the dataset while explicitly separating these measures from semantic correctness.
- Analyze typical error modes, uncertainty patterns, and practical limits of automated reconstruction for migration scenarios.

The main contributions are threefold:

- a provenance-aware intermediate data-architecture representation that assigns reconstructed elements an explicit epistemic class—declared, deterministically derived, or suggested—and retains available source or inference metadata;
- a deterministic-first reconstruction architecture in which LLM-generated hypotheses are isolated in a non-mutating enrichment overlay;
- an artifact-based evaluation procedure that separately measures execution behavior, structural preservation, and preliminary parent-table candidate admissibility without conflating these properties with semantic correctness.

## II. Related Work

### A. Database reverse engineering and conceptual schema recovery

Database reverse engineering (DBRE) focuses on recovering higher-level data structures from existing database implementations. This process extends beyond extracting tables, columns, and data types: it may also involve reconstructing entities, keys, relationships, constraints, and other elements of the conceptual design [1], [2], [13]. Related work also showed that higher-level entity–relationship structures can be identified from relational schemas by analyzing keys, inclusion dependencies, and structural patterns encoded in the physical design [18]. Hainaut et al. describe DBRE as a combination of data-structure extraction and data-structure conceptualization, in which implementation artifacts are transformed into progressively more abstract specifications [1].

A central difficulty is that a physical database schema may not provide a complete representation of its original conceptual model. Existing systems may have incomplete documentation, reorganized structures, performance-oriented modifications, or constraints that were never explicitly declared in the database [1], [13]. Consequently, reverse-engineering techniques cannot assume that every conceptual property has a direct representation in DDL. They must instead combine explicit schema declarations with information derived from names, data types, structural patterns, procedures, and other available evidence.

Hainaut et al. further emphasize that many earlier reverse-engineering approaches relied on unrealistic assumptions, including the availability of a complete physical schema, a direct translation from conceptual specifications, and consistently chosen names for related keys [1]. These limitations are particularly relevant to legacy migration projects, where schemas may have evolved incrementally and where the available artifacts may not reflect the original design decisions.

Alhajj proposed a method for extracting an extended entity–relationship model from a legacy relational database [2]. The approach addresses the recovery of entities, attributes, keys, and relationships when the original conceptual documentation is unavailable. This work provides a foundation for reconstructing logical and conceptual specifications from relational artifacts. However, recovered structures should be interpreted according to the strength of the evidence supporting them. A relationship explicitly declared by a foreign key is different from one inferred from column names or compatible data types.

Database reverse engineering can also be viewed as a process of transforming an implementation-level schema into a conceptual representation. Hainaut et al. present DBRE as a broader engineering activity that connects requirements, existing database structures, and computer-aided reengineering tools [13]. This perspective is useful for migration analysis because it treats reverse engineering not as a one-time extraction step, but as an organized process involving multiple representations and transformation activities.

Early DBRE systems already demonstrated that substantial parts of conceptual reconstruction can be automated, while also identifying cases in which human input remains necessary [8], [9]. Premerlani and Blaha proposed systematic relational-database reverse engineering into higher-level models [8], whereas Chiang, Barron, and Storey combined schema and instance analysis to recover a semantically richer EER representation [9]. Hainaut et al. subsequently focused on the elicitation of implicit structures, including relationships that are not directly represented by physical constraints [10], and Henrard and Hainaut showed that hidden data dependencies may require evidence obtained from program understanding [11]. More recent model-driven work represents DBRE as a chain of model-to-model transformations defined over explicit metamodels [12]. Together, these studies support a central premise of the present work: reconstructed architecture should preserve the distinction between directly observed constructs and elements recovered through additional reasoning.

### B. *Schema matching and heterogeneous database structures*

Schema matching addresses the problem of identifying correspondences between elements of heterogeneous schemas. Earlier work on database schema integration established that combining heterogeneous schemas requires explicit resolution of naming, structural, and semantic conflicts across different representations [19]. Rahm and Bernstein surveyed approaches that use names, data types, structural relationships, constraints, and instance-level information to establish matches between schemas [15]. Their analysis shows that schema matching is not a purely syntactic task: equivalent concepts may have different names, different structures, or different levels of granularity, while similar names may represent different meanings.

These observations are particularly important for cross-platform database migration. A direct mapping between source and target types or columns may preserve syntactic compatibility without preserving data semantics. Differences in type systems, constraints, key representations, and relationship structures can therefore require transformations at the logical or conceptual level. Schema matching research supports the use of multiple evidence sources and explicit correspondence representations rather than relying on literal name or type equality [15].

Subsequent schema-matching research has reinforced this multi-evidence view. Cupid combines linguistic similarity with data types, constraints, and schema structure [20], while Similarity Flooding represents schemas as graphs and propagates similarity through structural relationships [21]. COMA and COMA++ further demonstrate that no single matcher is sufficient across heterogeneous schemas and instead combine complementary matching strategies within configurable frameworks [22], [23]. Benchmarking work such as Valentine has also shown that matching quality depends substantially on the dataset characteristics and evaluation scenario, motivating explicit benchmark definitions rather than assuming that a matcher generalizes uniformly across schemas [24]. These findings support the present pipeline's use of structural catalog information as evidence while treating inferred correspondences as candidates rather than source facts.

Recent LLM-based schema-matching research extends these ideas by replacing or augmenting manually designed semantic similarity functions with language-model reasoning. Parciak et al. show that LLMs can identify semantic correspondences using schema names and descriptions alone, but also report that matching effectiveness is sensitive to the amount of contextual information supplied [25]. Retrieval-enhanced matching methods similarly use external or selectively retrieved schema context to constrain LLM reasoning rather than exposing an entire schema indiscriminately [26]. Magneto combines efficient candidate retrieval with LLM reranking, illustrating a staged architecture in which expensive semantic reasoning is applied only to a constrained candidate set [27]. These findings are consistent with the deterministic-first design adopted here: model reasoning is most defensible when applied after deterministic reconstruction has bounded the admissible schema space.

The amount and organization of schema context supplied to an LLM are themselves relevant design variables. Trummer demonstrates that relational-schema descriptions can be compressed substantially while preserving effectiveness for downstream LLM tasks, framing schema representation as an optimization problem rather than assuming that complete raw schema text is always desirable [28]. Cross-agent specification experiments further show that specification size alone does not predict downstream implementation quality and that a specification produced for one development agent may degrade substantially when consumed by another [29]. This motivates keeping the reconstructed specification explicit, structured, and independent of a particular model invocation rather than treating prompt text as the architecture representation itself.

The DBRE literature and schema-matching literature share a common concern: the information available in physical artifacts is often insufficient to determine intent uniquely. Several candidate relationships or mappings may be technically plausible. A reliable reconstruction process must therefore distinguish directly declared elements, deterministically derived elements, and model- or rule-generated hypotheses. The present study adopts this distinction by maintaining a declared graph separately from an enrichment overlay.

### C. *Evaluation, verification, and the evidence–hypothesis distinction*

Evaluation of reconstructed database structures requires more than measuring whether a pipeline completes or produces syntactically valid output. A reconstruction may be structurally plausible while containing incorrect keys, relationships, or domain interpretations. The DBRE literature treats conceptual recovery as an analytical process involving transformations and assumptions [1]–[3], [8]–[14]. Accordingly, evaluation should separately consider extraction behavior, output coverage, representation consistency, and semantic correctness.

This distinction is also relevant to the evaluation of LLM-generated artifacts. Wang et al. examined the use of LLMs as evaluators in software engineering and reported concerns about relying on model judgments as substitutes for independent validation [30]. The findings motivate caution when using an LLM to assess another model's inferred schema elements. A generated relationship should not be considered correct merely because it appears plausible to an automated evaluator.

The same principle appears in established schema-matching research, where automatically generated correspondences are commonly treated as candidate mappings that may require user verification rather than as unquestionable semantic facts [21]–[24]. This distinction becomes more important with generative models because fluent explanations and high model confidence are not independent evidence of relational validity. Consequently, candidate evaluation should preferably combine catalog-level checks, key and type compatibility, structural evidence, execution evidence where available, and independent human or reference-schema adjudication.

Instead, semantic validation should rely on known reference schemas, independent expert review, execution evidence, or other explicitly defined oracles. For this reason, the present study separates execution-oriented metrics from semantic-validation metrics. Pipeline completion, empty-result handling, failure behavior, key coverage, relationship coverage, and the availability of provenance metadata can be assessed directly from artifactized schema samples. In contrast, the semantic correctness of inferred primary-key and foreign-key candidates requires an independently reviewed reference or equivalent validation process. This separation prevents successful execution of the reconstruction pipeline from being confused with correctness of every inferred architectural element.

### D. Provenance and Schema Evolution

Database reconstruction is also closely related to schema evolution and provenance. Database evolution research has shown that schema transformations affect not only the structural representation of data but also queries, applications, documentation, and the ability to explain how derived artifacts were obtained [31]. Provenance research formalizes this requirement by distinguishing the origin and derivation of data products and by preserving information about how outputs depend on prior representations [32], [33]. For reverse engineering, the same principle applies at the schema level: an architectural element recovered directly from DDL has a different evidential status from one introduced by a deterministic transformation or inferred from contextual evidence.

Curino et al. demonstrated that automated schema evolution can combine migration, query rewriting, documentation, historical access, and provenance rather than treating schema conversion as an isolated rewrite [30]. This perspective motivates maintaining provenance across reconstruction stages. In the present work, provenance is therefore attached to reconstructed elements and retained across the catalog, graph, enrichment overlay, and generated specifications. The declared/derived/suggested partition operationalizes this distinction by preventing later transformations from erasing how a particular architectural statement entered the reconstructed model.

### E. Research gap

Existing DBRE research provides mature techniques for recovering conceptual structures, dependencies, keys, cardinalities, and other semantics from relational implementations [1]–[3], [8]–[14], [18]. Cardinality reconstruction from relational constraints has also been studied explicitly in DBRE, reinforcing the need to distinguish declared relationship structure from properties inferred from available constraints [34]. Schema-matching research provides complementary methods for establishing correspondences through names, types, constraints, graph structure, instance evidence, and combinations of multiple matchers [15], [20]–[24]. More recently, LLM-based schema-matching studies have demonstrated that language models can contribute semantic evidence, but they also show sensitivity to context selection, computational cost, and candidate-space construction [25]–[27].

A separate body of specification-driven research treats specifications as structured operational artifacts whose quality depends on explicit context, traceability, determinacy, portability, and treatment of unresolved information [5]–[7], [29]. However, these research lines remain only partially connected. Traditional DBRE approaches generally do not model LLM-generated hypotheses as a separately governed enrichment layer; LLM schema-matching approaches typically seek correspondences rather than reconstructing a complete provenance-preserving logical and conceptual architecture from incomplete physical DDL; and specification-driven approaches normally assume that a usable specification already exists rather than reconstructing it from implementation evidence. Schema-evolution and provenance research additionally motivates preservation of derivation history [31]–[33], but does not by itself provide a deterministic-first LLM-assisted reconstruction process.

The specific gap addressed here is therefore the absence of an integrated process that combines deterministic physical-schema extraction, rule-governed recovery, declared relationship-graph construction, model-assisted hypothesis generation, abstraction into logical and conceptual specifications, and element-level provenance while keeping source-grounded and deterministically derived evidence distinct from unresolved or model-suggested architecture.

The present study addresses these gaps through a staged reconstruction process for Oracle-oriented physical schemas. It combines deterministic extraction and recovery with model-assisted enrichment, maintains declared and inferred graph elements separately, records provenance and confidence for suggestions, and evaluates both execution behavior and output structure. The resulting process treats reconstruction as a controlled transformation from physical evidence to explicit intermediate specifications rather than as an unverified generation task.

## III. Method and Experimental Design

### A. Problem formulation

We formulate database reconstruction as the transformation of a set of physical DDL artifacts into a structured, partially observed data-architecture specification.

Let the source database schema be a finite set of relation schemas:

$$DB = \{R_1, R_2, ..., R_n\}. \quad (1)$$

Each relation schema is represented as:

$$R = (N, A, dom, C), \quad (2)$$

where $N$ is the relation name, $A$ is the set of attributes, $dom$ is a function assigning a value domain to each attribute, and $C$ is a tuple of integrity constraints. For a relation $R$, the attribute set is:

$$A = \{a_1, a_2, ..., a_m\}, \quad (3)$$

and the domain function is $dom(a_i) = d_i$, where $d_i$ includes the type and relevant physical parameters of attribute $a_i$, such as length, precision, scale, or temporal properties.

We represent the constraint tuple as:

$$C = (PK, FK, UQ, NN, CHECK), \quad (4)$$

where $PK \subseteq A$ is the primary-key attribute set, $UQ \subseteq 2^A$ is the set of unique attribute subsets, $NN \subseteq A$ is the set of non-nullable attributes, $FK$ is the set of references from attributes in the current relation to attributes of other relations and $CHECK$ is the set of predicates over relation tuples. A foreign-key element can be written as:

$$(A_c, R_p, A_p), \quad (5)$$

where $A_c \subseteq A$ is the referencing attribute set, $R_p \in DB$ is the referenced relation, and $A_p \subseteq A(R_p)$ is the referenced attribute set. The physical schema may also contain platform-specific declarations, defaults, generated-column expressions, and type parameters. These are retained as source attributes or mapped to general logical type categories where a deterministic mapping is available.

The input to reconstruction is not $DB$ directly but a collection of DDL artifacts,

$$S = \{s_1, s_2, ..., s_k\}, \quad (6)$$

where each artifact may contain table definitions, constraints, alterations, indexes, or statements that are irrelevant to the target schema. The parser produces observations from $S$, and the consolidation stage normalizes identifiers, merges repeated table definitions, extracts declared constraints, and records source-file links. Consequently, the observed catalog is a partial and potentially inconsistent view of $DB$ rather than a complete oracle.

To make this distinction explicit, we partition reconstructed information into three classes:

$$C_{obs} = C_{declared} \cup C_{derived} \cup C_{suggested}, \quad (7)$$

where the three classes are mutually exclusive with respect to the provenance status assigned to each reconstructed element.

$C_{declared}$ contains elements explicitly recovered from source DDL, including primary-key and foreign-key declarations recovered from supported CREATE TABLE and ALTER TABLE statements. $C_{derived}$ contains elements produced from source-grounded evidence by deterministic rules, including identity-column or naming heuristics, type normalization, and modality or cardinality derivation from nullability, uniqueness, and key information. $C_{suggested}$ contains model-assisted candidates for missing primary keys or foreign-key relationships.

The partition in (7) is epistemic rather than merely structural: these classes are not interchangeable. A suggested relationship is a hypothesis about the latent schema, not evidence that the relationship is declared in the source. This partition also corresponds to a broader distinction between specification evidence and specification completion. Formal specification work argues that unresolved decisions should remain explicitly open when available evidence does not determine a unique implementation [7].

Accordingly, $C_{suggested}$ is not treated as an extension of $C_{declared}$ or $C_{derived}$; it defines a separately governed hypothesis layer whose elements require independent evidence before promotion to the reconstructed baseline. The representation therefore preserves epistemic status together with structural content.

Using the provenance partition in (7), the reconstruction target is defined by (8) as an intermediate specification:

$$\widehat{DB} = (\hat{R}, G, P), \quad (8)$$

where $\hat{R}$ is the reconstructed relation catalog, $G$ is the declared relationship graph, and $P$ is provenance metadata. For each reconstructed element $x$, $P(x)$ records its source class, and, when applicable, source file, confidence, and inference rationale. Missing elements are represented as missing or as candidates rather than silently materialized as declared structure.

The task is not to recover a unique true conceptual model from syntax alone. Instead, given $S$, the pipeline computes a traceable approximation:

$$f(S) = (\hat{DB}), \quad (9)$$

such that directly observed facts remain linked to their DDL evidence, deterministic transformations are distinguishable from observations, and uncertain model-assisted additions remain reviewable.

Evaluation consequently separates pipeline and representation properties from semantic validity. Completion and output coverage can be measured directly from generated artifacts; correctness of inferred keys, foreign keys, and conceptual relationships requires an independent reference or expert review and is not assumed by the formalization.

The experimental question is whether the proposed reconstruction process can:

- complete successfully on heterogeneous Oracle-oriented DDL inputs;
- preserve deterministically recoverable table and column structure across abstraction levels;
- retain provenance and unresolved information without contaminating the declared baseline;
- expose the admissibility limits of LLM-generated PK and FK hypotheses.

Semantic correctness is outside the validated scope of the present experiment because no independently annotated reference schema is available.

*B. End-to-End pipeline*

The reconstruction pipeline is organized as a deterministic-first sequence followed by an optional model-assisted enrichment stage. Its input is an ordered list of SQL files belonging to one schema sample. Each stage consumes a typed result from the preceding stage and writes inspectable artifacts, allowing intermediate failures and information loss to be localized rather than observed only at the final specification. The complete stage sequence, including the separate enrichment path and stage-specific error artifacts, is shown in Fig. 1.

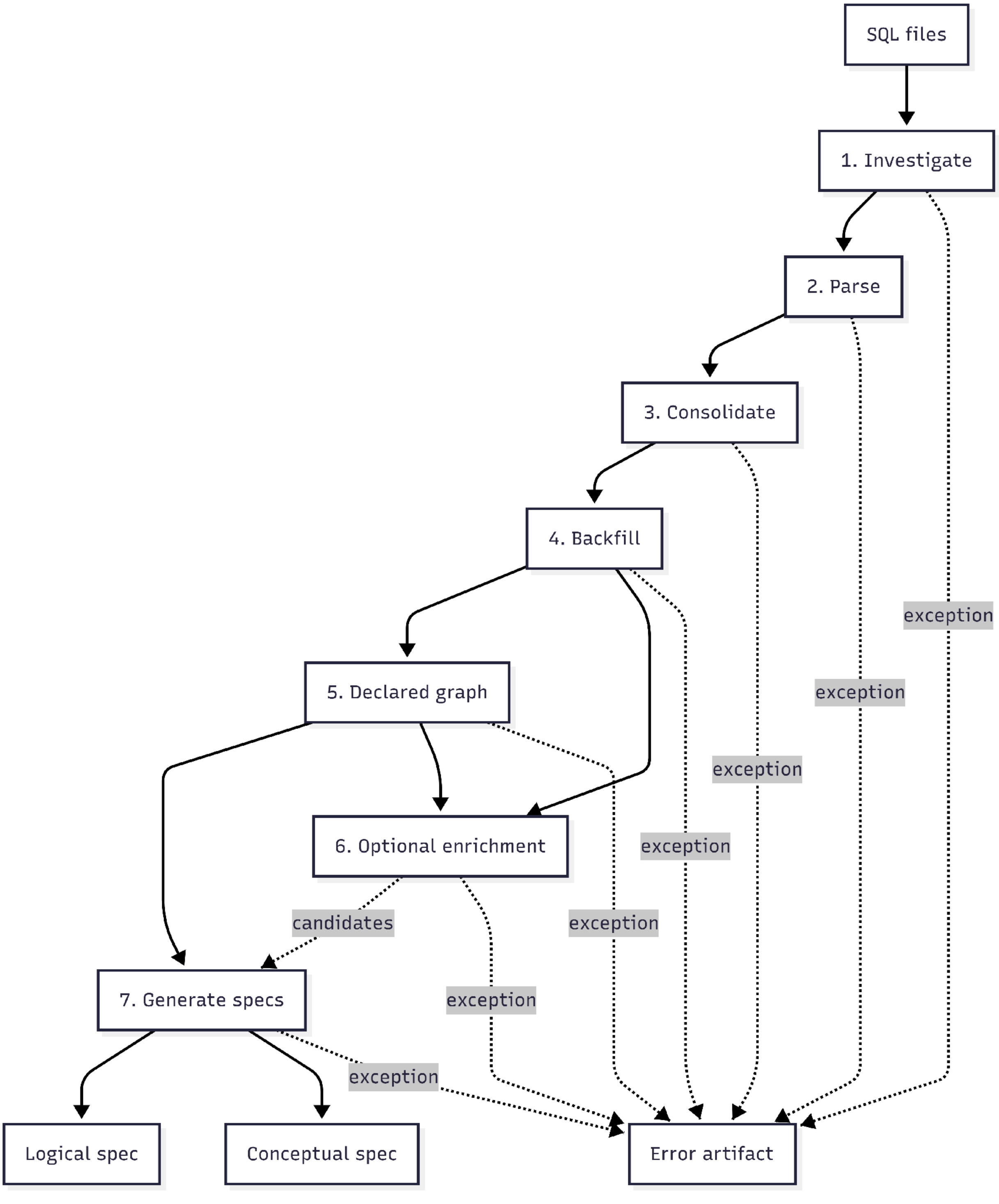

Fig. 1. Deterministic-first pipeline for physical-to-logical/conceptual data-architecture reconstruction.

As Fig. 1 illustrates, deterministic stages construct the source-grounded baseline before optional enrichment, while failures from individual stages are preserved as explicit error artifacts.

*C. Investigation and DDL isolation*

The investigation stage reads each input file using UTF-8 with a Latin-1 fallback, records the original file order, and extracts table-creation and table-alteration statements. Statements associated with data manipulation or transaction control are flagged and excluded from the combined DDL stream. The stage produces the combined DDL text and a per-file inventory containing file paths, ordering, blocked-keyword hits, and extracted text lengths. This filtering bounds the parser input to schema-definition statements while retaining enough metadata to audit which files contributed evidence.

### D. Physical DDL parsing

Each DDL fragment is parsed independently with a dedicated DDL parsing library configured for Oracle syntax. Parser results are retained per file and reduced to table records containing the table name, columns, physical types and attributes, explicitly declared primary keys, and constraint information. Files with no extracted DDL produce an empty parse result rather than an artificial table. Parse statistics record the number of files, tables, tables per file, and tables without an explicitly parsed primary key. Parser output is treated as an observation layer: an absent field means that the parser did not return that information, not necessarily that the source database lacks the corresponding construct.

### E. Consolidation and normalization

The consolidation stage merges table records across files into a global catalog. Identifiers are normalized by removing quoting and applying consistent case handling; missing schemas are assigned a common unknown-schema value. A table record contains its normalized schema-qualified name, columns, declared primary key, contributing source files, and raw parser sources. Foreign keys are extracted from column-level references, table constraint references, and supported table-alteration structures. Duplicate foreign-key records are removed using their child relation, child columns, parent relation, and parent columns as a signature. Conflicting table or column definitions are retained as catalog issues with source-file references rather than silently overwritten.

### F. Deterministic backfill and type normalization

Backfill completes selected information that can be derived from source text or local structural rules. First, primary keys declared in supported table-alteration statements are recovered with a pattern-based pass and applied only when they do not conflict with an existing declaration. Tables still lacking a primary key are examined using identity-column and naming heuristics. These heuristic results are recorded as derived information and associated with an issue record. In parallel, Oracle physical types are mapped to general logical categories such as character, integer, decimal, date-time, binary, and identifier types, while preserving the original physical type and mapping notes. For declared foreign keys, modality and cardinality are derived from nullability, uniqueness, and key information, with confidence fields recording the status of the derivation.

### G. Declared relationship graph

The graph stage constructs a directed schema graph from the consolidated catalog after backfill. Each node represents a normalized, schema-qualified table and stores basic table metadata, including column count and available primary-key information. Each declared foreign key contributes an edge from the child table to the parent table, carrying child and parent columns, constraint name, source kind, source file, modality, cardinality, and their confidence values. Referenced tables that are not present as complete table records are retained as nodes marked as referenced-only. The resulting graph is intentionally declared-only; it is not mutated by subsequent model inference.

### H. Model-assisted enrichment overlay

The enrichment stage receives the backfilled catalog and the declared graph as read-only context. When enabled, it requests primary-key candidates for tables that still lack a deterministic primary key and foreign-key candidates for tables without outgoing declared edges. Candidate generation is conditioned on the reconstructed table catalog, but model outputs are not guaranteed to reference only catalog-resident parent tables; this condition is therefore checked explicitly during evaluation.

This conditioning follows the candidate-generation principle used in staged schema-matching architectures: model reasoning should operate over a bounded candidate space constructed from reliable structural evidence rather than generate arbitrary correspondences over an unconstrained vocabulary [27]. Parent-table membership in the reconstructed catalog is treated as a preliminary structural admissibility condition verified explicitly during evaluation, rather than as a semantic validation rule. A candidate that satisfies it may still be invalid because of incompatible columns, key membership, direction, modality, or domain meaning.

For every candidate, the overlay stores the source and target nodes, child columns, model confidence, and model rationale. Modality and cardinality for suggested foreign keys are derived using the same local catalog information used for declared relationships, but the edge remains a suggestion and is not inserted into the declared graph. Configuration flags allow primary-key and foreign-key inference to be disabled independently, and the model configuration is supplied separately from deterministic stages.

### I. Logical and conceptual specification generation

The specification stage combines the backfilled catalog, the declared graph, and the enrichment overlay into two outputs. The logical specification represents tables as entities with attributes, mapped logical types, nullability and uniqueness information, primary-key information, and relationships. Declared relationships retain source provenance, whereas model-assisted primary-key and foreign-key additions are labeled as suggested and retain confidence or rationale metadata. The conceptual builder uses the same inputs to generate business-oriented entity names, relationship descriptions, and business-rule candidates through model-assisted calls. Both specifications and their aggregate statistics are serialized separately from the intermediate artifacts.

Related architecture-metamodel research likewise motivates preserving multiple architectural views within a common traceable representation rather than reducing system knowledge to a single generated artifact [35]. In an LLM-mediated workflow, this separation additionally allows each abstraction to be supplied, reviewed, or regenerated independently and reduces dependence on a particular model-specific prompt representation [5], [29].

For each sample, the orchestrator writes artifacts for the investigation, parse, consolidation, backfill, graph, enrichment, and specification stages. A stage exception is recorded in a dedicated error artifact together with the ordered input files and traceback before the exception is propagated. This execution model supports two complementary analyses: deterministic intermediate artifacts can be inspected for completeness and consistency, while enrichment and specification artifacts can be reviewed for the plausibility and provenance of model-assisted additions.

### J. Dataset Construction and Characteristics

The primary dataset consists of Oracle-oriented DDL artifacts organized as an ordered collection of files for each schema sample. The dataset loader groups files by sample identifier and orders the files within each group by their sequence number. This preserves the multi-file structure and statement order supplied to the orchestration pipeline. The available corpus contains 1,225 SQL files grouped into 249 samples. Each sample is treated as an independent experimental unit. Only schema-level SQL artifacts are used; database instance rows, application source code, runtime traces, and external domain documentation are not supplied to the reconstruction pipeline.

The corpus is an operationally heterogeneous evaluation corpus rather than a controlled synthetic benchmark; consequently, the experiment evaluates pipeline robustness and structural preservation under naturally varying DDL composition, not statistical representativeness of the broader Oracle ecosystem. A sample may contain one or more SQL files, table-creation statements, subsequent table-alteration statements, and statements that are filtered during investigation.

Consequently, the experimental inputs expose the pipeline to incomplete declarations, parser edge cases, repeated definitions, ALTER-based constraints, and cross-file ordering dependencies while preventing instance-level data from providing additional semantic evidence. It also means that an empty output is a meaningful execution category: it may reflect a sample without extractable table DDL or a limitation of the current extraction and parsing path. The dataset is therefore used to assess robustness and coverage, not assumed to provide complete ground-truth logical or conceptual models.

For the current artifactized run, 244 samples completed the orchestration stages and five produced failure artifacts. The completed samples contain the intermediate catalogs, declared graphs, enrichment overlays, logical and conceptual specifications, and downstream evaluation artifacts. These counts describe execution availability; they do not imply that the reconstructed keys, relationships, or conceptual interpretations are correct.

### K. Artifact Taxonomy and Representation

The pipeline uses a staged artifact taxonomy to preserve both transformation state and provenance. Table I summarizes the artifacts produced at each reconstruction stage and the information retained for subsequent analysis and provenance tracking.

Investigation artifacts contain the combined DDL and the contributing-file inventory. Parse artifacts contain per-file parser outputs and parse statistics. Consolidation artifacts contain the normalized table catalog, declared foreign keys, source-file references, and catalog issues. Backfill artifacts contain the post-completion catalog and statistics for recovered keys, remaining key gaps, type mapping, and derived relationship properties. Graph artifacts serialize the declared schema graph and its edge list.

TABLE I. STAGE ARTIFACT TAXONOMY OF THE RECONSTRUCTION PIPELINE.

| Stage | Artifact contents |
|---|---|
| Investigation | Combined DDL text; per-file inventory (paths, ordering, blocked-keyword hits, extracted text lengths) |
| Parse | Per-file parser output; parse statistics (files, tables, tables/file, tables without parsed PK) |
| Consolidation | Normalized table catalog; declared FKs; source-file references; catalog issues |
| Backfill | Post-completion catalog; recovered-key statistics; remaining key gaps; type-mapping notes; derived relationship properties |
| Graph | Declared schema graph (nodes, edges); edge list with child/parent columns, constraint name, modality, cardinality, confidence |
| Enrichment | PK/FK candidate overlay (source, target, columns, confidence, rationale); not merged into declared graph |
| Specification | Logical spec (entities, attributes, types, keys, relationships); conceptual spec (business names, relationship sentences, business-rule candidates); aggregate statistics |

Enrichment artifacts are kept separate from the declared graph. The LLM overlay records primary-key candidates and foreign-key suggestions, including confidence and rationale; it does not replace the declared catalog or mutate the declared graph. Specification artifacts then expose two abstraction levels. The logical specification contains schema-qualified entities, attributes, logical type objects, nullability, uniqueness, primary-key sources, and declared or suggested relationships. The conceptual specification contains LLM-derived entity names, relationship sentences, business-rule statements, and technical links back to the underlying tables and constraints.

For downstream round-trip analysis, each completed sample may additionally contain four generated Oracle DDL variants: logical without LLM additions, logical with LLM additions, conceptual without LLM additions, and conceptual with LLM additions. Mermaid ER artifacts are generated for the original catalog and for logical and conceptual specifications with and without LLM additions. This taxonomy makes it possible to compare source-derived structure, reconstructed specifications, and generated implementation artifacts without collapsing them into one undifferentiated output.

### L. Provenance-Aware Intermediate Data-Architecture Model

The provenance-aware intermediate data-architecture model is the structured hand-off between physical DDL analysis and logical/conceptual specification generation. Its physical-facing component is the consolidated catalog: normalized schema and table identifiers, column records, physical types, sizes, nullability, uniqueness, defaults, checks, generated-column information, declared primary keys, source files, and recorded issues. Its relationship component is a collection of foreign-key records containing child and parent relations and columns, constraint names, source kind, source file, modality, cardinality, and confidence fields.

The model also exposes a graph view in which schema-qualified tables are nodes and declared foreign keys are directed edges from child to parent. Nodes may be marked as present or referenced-only when a relationship names a table not represented by a complete table definition. The logical specification projects this information into technology-independent entity and relationship objects while retaining physical-type information and mapping notes. The conceptual specification projects the same evidence into business-oriented names, natural-language relationship descriptions, and business-rule candidates. The model is consequently an intermediate representation rather than a claim that the source schema has been fully recovered: source class, provenance, confidence, and missingness remain part of the representation.

Fig. 2 summarizes the provenance-aware intermediate representation and shows how source artifacts, deterministic reconstruction, derived information, and model-assisted suggestions remain connected while preserving their distinct evidence status.

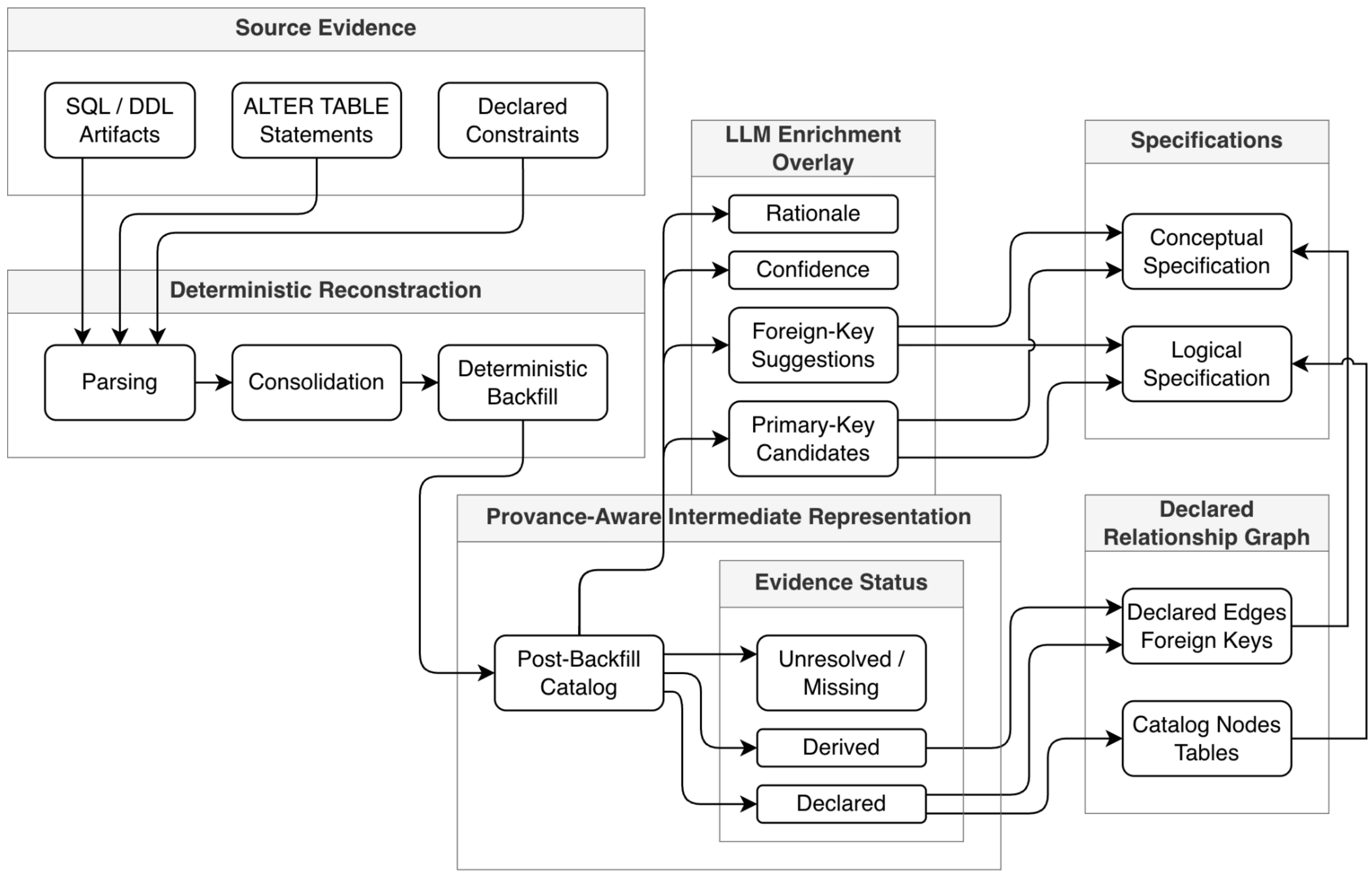

**Fig. 2.** Provenance-aware intermediate representation linking source evidence, deterministic reconstruction, derived information, and model-assisted specification artifacts.

## *M. LLM Configuration and Credential Handling*

Model-assisted stages use an Azure OpenAI endpoint configured externally through environment variables. The experimental configuration records the model identifier, API version, endpoint configuration, and sampling temperature; temperature is fixed at zero for the reported run. Credentials are excluded from generated artifacts.

Reproduction of LLM-enabled stages therefore requires the operator to provision valid credentials outside the repository and expose them through the expected environment variables. Runs without those variables fail during LLM-client construction. Deterministic stages can be analyzed independently of model credentials, while reported LLM-dependent results must be interpreted together with the model identifier, API version, temperature, and credentialed execution environment.

## *N. Implementation and Reproducibility Details*

The implementation is a Python pipeline organized into investigation, parsing, consolidation, backfill, graph, enrichment, and specification modules. A dataset-execution component discovers and orders the sample files, creates a separate artifact namespace for each sample, invokes the orchestration runner, skips samples with existing outputs by default, and records failures. The orchestration runner writes numbered JSON, SQL, CSV, and graph artifacts after each completed stage, and writes an error record before propagating an exception.

The execution procedure is reproducible at the pipeline level through the documented command-line interface, input ordering rules, stage artifacts, and fixed temperature setting; exact LLM outputs remain dependent on the externally hosted model version and service state. Individual samples can also be executed by supplying their numeric identifiers to the sample-level execution function. A separate evaluation component reads completed sample artifacts, generates the four configured DDL variants, computes round-trip metrics, writes per-sample metric records, and emits Mermaid ER representations. The principal evaluation configuration uses a sampling temperature of zero.

The evaluation normalizes the backfilled catalog as the reference representation and compares it with normalized logical or conceptual outputs. Metrics include table preservation, column preservation, physical-type agreement, primary-key agreement, declared foreign-key counts, unresolved-PK incidence, and LLM candidate admissibility. Table and column preservation measure structural carry-through relative to the post-backfill catalog; type agreement measures normalized physical-type consistency; candidate admissibility measures whether an LLM-suggested parent table exists in the reconstructed reference catalog.

## IV. Results and Discussion

### A. Evaluation Basis

Table II summarizes the evaluation basis. The pipeline was executed on 249 Oracle-oriented DDL samples comprising 1,225 SQL files. Each sample was processed through investigation, parsing, consolidation, deterministic backfill, declared-graph construction, optional enrichment, and specification generation.

TABLE II. Evaluation Basis

| Component | Basis |
|---|---|
| Dataset | 249 samples; 1,225 SQL files |
| Completed-sample denominator | 244 samples |
| Reference representation | Post-backfill catalog |
| Output variants | Logical and conceptual, each with and without LLM additions |
| Enrichment configuration | PK and FK inference enabled |
| Preserved artifacts | Stage JSON, SQL, CSV, graph, and error artifacts |

The round-trip evaluation compares each generated specification with the post-backfill catalog, which is the deterministic reference produced before LLM intervention. Unless stated otherwise, sample-level metrics are macro-averaged over the 244 completed samples; aggregate counts such as reconstructed tables and declared foreign keys are corpus-level totals and should not be interpreted as macro-averages. This design measures execution, structural preservation, and agreement with the reconstructed reference. It does not provide an independently annotated semantic oracle for primary keys, foreign keys, or conceptual interpretations.

The resulting evaluation boundary is illustrated in Fig. 3, which separates properties established directly from pipeline artifacts from claims that require independent schema or domain validation.

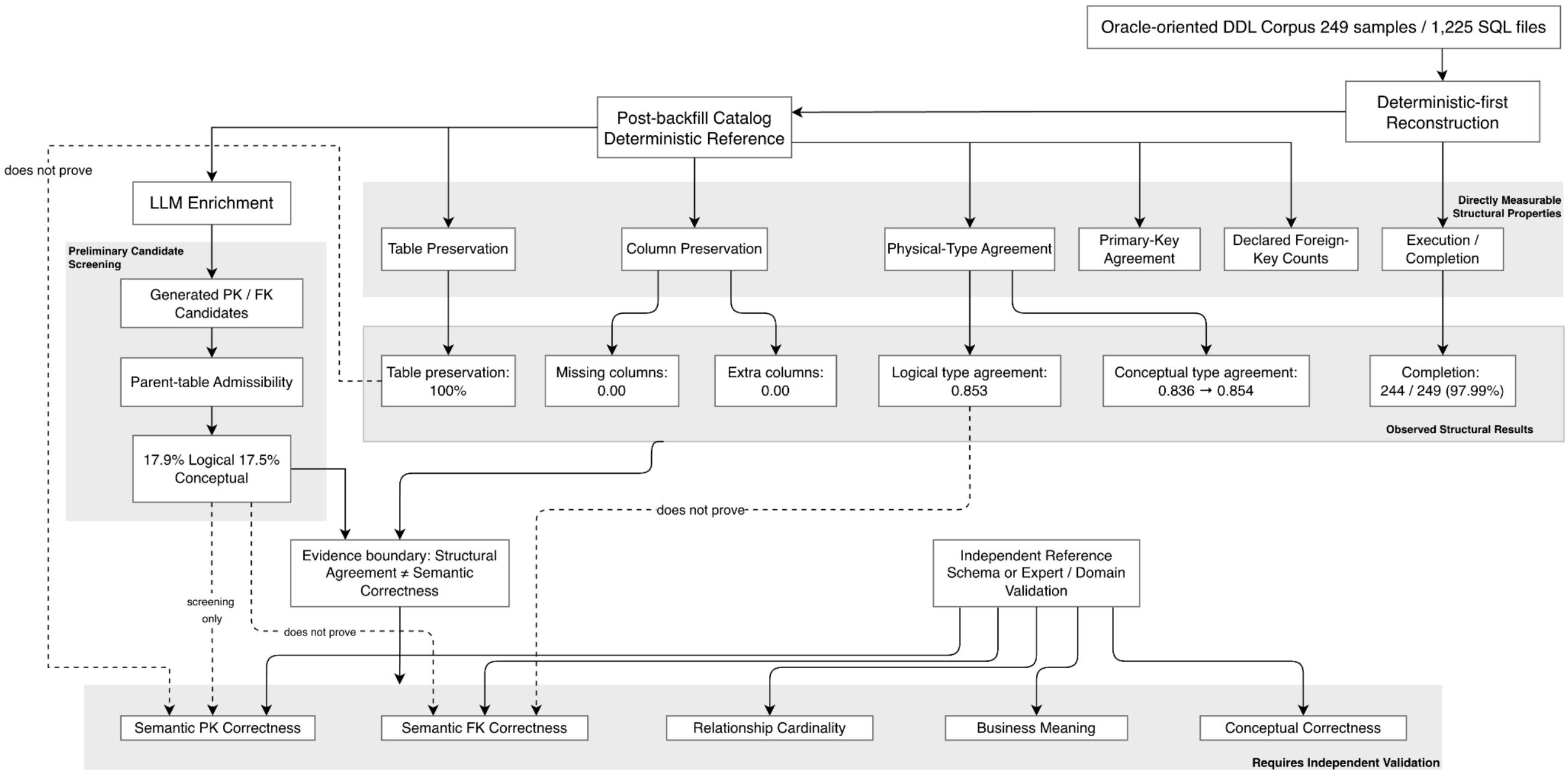

**Fig. 3.** Evaluation and evidence boundary separating directly measurable reconstruction properties from semantic claims requiring independent validation.

As Fig. 3 shows, the current evaluation establishes execution behavior, structural preservation, type agreement, and preliminary candidate admissibility, but it does not establish semantic correctness of inferred keys, relationships, cardinalities, business meaning, or conceptual interpretations. These properties remain outside the validated evidence boundary and require an independent reference schema or expert/domain validation.

### B. Execution, Extraction, and Deterministic Baseline

Table III summarizes execution outcomes. The pipeline completed for 244 of 249 samples (97.99%). The five failures were concentrated in parser-library edge cases involving ALTER TABLE variants. Empty outputs were retained as an explicit outcome when investigation found no extractable CREATE TABLE or ALTER TABLE statement; they were not converted into artificial schema structure.

TABLE III. PIPELINE EXECUTION OUTCOMES OVER 249 INPUT SAMPLES.

| Outcome | Samples | Share of all samples |
|---|---|---|
| Completed | 244 | 97.99% |
| Failed | 5 | 2.01% |
| Completed with no extractable DDL | 52 | 20.88% |
| Completed with a non-empty reconstructed schema | 192 | 77.11% |

In the reported experimental configuration, enrichment was enabled for all 244 completed samples; therefore, enrichment-stage availability was 100% among completed samples and 97.99% over the full corpus. The investigation stage preserved file order and recorded source-file inventories, blocked-keyword observations, and extracted text lengths.

Consequently, the execution results distinguish three operational outcomes—successful reconstruction, successful execution with no extractable schema structure, and pipeline failure—rather than collapsing empty outputs and failures into a single unsuccessful category.

The completed deterministic baseline contained 208 reconstructed table definitions and 278 declared foreign-key records. At parse time, 168 of the 208 tables lacked an explicitly parsed primary key. This absence is an observation about parser output, not proof that the source schema has no primary-key declaration.

Consolidation normalized identifiers, retained catalog issues instead of overwriting conflicts, and deduplicated foreign keys by child relation and columns, parent relation and columns. The declared graph therefore has 208 catalog nodes and 278 declared edges; referenced-only parents are preserved where encountered. Table IV summarizes the resulting deterministic extraction and primary-key backfill outcomes.

TABLE IV. DETERMINISTIC EXTRACTION AND BACKFILL OUTCOMES.

| Deterministic artifact | Aggregate result | Interpretation |
|---|---|---|
| Reconstructed tables | 208 | Corpus-level total across 192 non-empty completed samples |
| Tables without an explicitly parsed PK | 168 | 80.8% of reconstructed tables at parse time |
| Declared foreign-key records | 278 | Corpus-level total across reconstructed schemas |
| Non-empty samples with no remaining PK gap after backfill | 29 | 15.10% of non-empty completed samples |
| Non-empty samples with at least one PK gap after backfill | 163 | 84.90% of non-empty completed samples |

The main deterministic-reconstruction finding is that primary-key incompleteness remained substantial: 163 of the 192 non-empty completed samples (84.90%) retained at least one unresolved PK gap after backfill. Thus, deterministic reconstruction provides a traceable baseline but does not eliminate the need for subsequent review or hypothesis generation. Backfill recovered supported ALTER TABLE primary-key declarations when non-conflicting and retained them as declared information because they are explicitly present in the source DDL. Identity-column and naming-heuristic results are retained as derived information with issue records. Oracle physical types were mapped to general logical categories while preserving the original type and mapping notes. For declared foreign keys, modality and cardinality were derived from nullability, uniqueness, and key information and stored with confidence metadata. These operations enrich the deterministic representation but do not change the evidence status of an element into a source declaration.

*C. Structural Preservation and Representation Variants*

Across the 192 completed samples with non-empty reconstructed schemas, each evaluated logical and conceptual representation preserved all 208 table records contained in the internal post-backfill reference.

Across all four variants, average missing and extra columns were both 0.00 per sample. This establishes exact table- and column-inventory preservation relative to the internal post-backfill reference for the evaluated outputs, but it does not establish semantic correctness of types, keys, or relationships. Table V summarizes structural preservation and type agreement across the four evaluated representation variants.

TABLE V. STRUCTURAL PRESERVATION AND TYPE AGREEMENT ACROSS REPRESENTATION VARIANTS.

| Variant | Table preservation | Avg. missing/extra columns | Type agreement |
|---|---|---|---|
| Logical, no LLM | 208/208 | 0.00 | 0.853 |
| Logical, with LLM | 208/208 | 0.00 | 0.853 |
| Conceptual, no LLM | 208/208 | 0.00 | 0.836 |
| Conceptual, with LLM | 208/208 | 0.00 | 0.854 |

The results show complete table and column inventory preservation across all four variants, while the observed differences are confined to type agreement rather than structural coverage.

The logical specification achieved a physical-type agreement of 0.853 both without and with LLM additions. Relative to the internal post-backfill reference, the conceptual-specification variants achieved normalized type-agreement values of 0.836 without LLM additions and 0.854 with LLM additions. These results indicate that both abstraction levels preserve table and column inventories, while type agreement remains sensitive to the abstraction process. The observed 0.018 increase for the conceptual variant after enrichment is descriptive only and cannot be attributed causally to LLM enrichment without paired statistical validation.

*D. LLM Enrichment: Coverage and Screening Results*

Enrichment generated primary-key candidates for samples with unresolved deterministic key gaps and foreign-key candidates for tables without declared outgoing relationships. Suggestions were stored in a separate overlay with confidence and rationale; the declared graph was not modified. This separation is central because the observed candidate coverage does not validate the candidates semantically.

The 100 foreign-key candidates were generated in 36 samples. A candidate passed the reported screening check only when its named parent table was present in the reference catalog. The parent-table admissibility rates of 17.9% for the logical variant and 17.5% for the conceptual variant imply that most suggested targets did not match a reconstructed reference table. Table VI summarizes the scope of LLM enrichment, the remaining primary-key gaps after deterministic backfill, and the parent-table admissibility observed for generated foreign-key candidates.

TABLE VI. LLM ENRICHMENT COVERAGE AND PARENT-TABLE ADMISSIBILITY.

| **Metric** | **Value** |
|---|---|
| Samples with enrichment enabled | 244 (100%) |
| Non-empty completed samples | 192 |
| Samples with ≥1 unresolved PK gap | 163 (84.90%) |
| Samples with no PK gap after backfill | 29 (15.10%) |
| FK candidates generated | 100 |
| Samples contributing FK candidates | 36 |
| Admissibility — logical variant | 17.9% |
| Admissibility — conceptual variant | 17.5% |

The results indicate that LLM enrichment primarily operates in areas where deterministic reconstruction remains incomplete, but the low parent-table admissibility values confirm that generated relationships should remain reviewable hypotheses rather than declared schema elements. This result empirically supports the central architectural decision to keep model-generated relationships outside the declared graph: even the preliminary parent-table existence condition is satisfied by only 17.5–17.9% of evaluated candidates, so automatic promotion of such hypotheses would risk contaminating the source-grounded baseline.

This result is also consistent with recent evidence that LLM-based schema matching depends strongly on how candidate context is constructed and that LLM-based semantic reasoning operating under only limited structural constraints should not be assumed to provide reliable correspondences by itself [25], [27]. In contrast to systems that rank predefined source-target candidate pairs, the present enrichment stage generates missing-relationship hypotheses; the observed low parent-table admissibility rate therefore reinforces the need for a tighter retrieval or candidate-generation layer before model reasoning. A promising extension would first retrieve structurally admissible parent-table candidates using names, type compatibility, key structure, and graph context and only then ask the LLM to rank or explain those candidates.

### *E. Interpretation and Implications*

The deterministic-first architecture succeeded in delivering a high-completion, auditable structural baseline: 244 of 249 samples completed (97.99%), yielding 208 reconstructed tables, 278 declared foreign-key records, explicit source links, and per-stage error artifacts that enable failure localization. For the evaluated corpus, this architecture supports three concrete pre-migration activities: generation of an auditable schema inventory, localization of unresolved key and relationship gaps, and prioritization of human review without modifying the source-grounded baseline. However, the enrichment results reveal why separation of declared and model-suggested elements is operationally essential. Among the 192 completed samples with a non-empty reconstructed schema, 163 (84.90%) retained at least one unresolved primary-key gap after deterministic backfill. In the reported enrichment configuration, the LLM stage generated 100 foreign-key candidates across 36 samples, with parent-table admissibility rates of 17.9% for the logical variant and 17.5% for the conceptual variant. The low parent-table admissibility rate observed under this preliminary screening criterion, that a suggested parent table exists in the recovered schema, demonstrates that model output should be used to prioritize human review rather than to materialize candidates as declared relationships. The principal architectural contribution is the explicit preservation of epistemic status across reconstruction stages: deterministic evidence forms the source-grounded baseline, whereas LLM-generated hypotheses are confined to a non-mutating enrichment overlay. This preserves the provenance class of baseline elements during enrichment: model-generated hypotheses may augment review context but cannot silently mutate the source-grounded architectural baseline.

Overall, the experiment yields three principal findings: the deterministic pipeline achieves a high execution-completion rate on the evaluated corpus; structural table/column information is preserved through logical and conceptual reconstruction; and LLM-generated relationship hypotheses require explicit screening and must remain separated from source-grounded architecture.

The choice between logical and conceptual outputs should reflect downstream purpose, not an assumption of semantic superiority. The logical variant preserves a physical-type agreement of 0.853 without enrichment and maintains that rate after enrichment. The conceptual variant begins at 0.836 (0.017 lower due to the additional abstraction layer) but recovers to 0.854 after enrichment. Both retain all 208 table records from the internal post-backfill reference and show 0.00 average missing or extra columns, establishing structural carry-through. For tasks prioritizing direct structural translation from physical DDL to a platform-independent logical form, the logical representation is the more direct choice. For communication of conceptual intent and business-oriented specification, the conceptual output supports human review, though semantic validation still requires that the reference itself be independently verified rather than circular confirmation against a reconstructed post-backfill catalog.

Several constraints temper the strength of these findings. Sample-level metrics are macro-averaged over the applicable completed samples, whereas reconstructed-table, key, relationship, and candidate counts are corpus-level aggregates. The corpus is heterogeneous, and samples of substantially different schema sizes therefore contribute equally to macro-averaged measures. No confidence intervals, per-sample distributions, or statistical significance tests were computed; differences such as the 0.017 physical-type-agreement difference between logical and conceptual baselines should not be interpreted as established without paired comparisons. The five failed samples may represent systematically harder cases; if excluded, completion rates may overestimate robustness. Parser coverage is bounded by supported Oracle DDL syntax, and LLM-dependent results are tied to the specific Azure OpenAI service, model version, API version, temperature, and execution context documented in Section III-N. The round-trip evaluation itself uses model-assisted judgment, creating a risk of correlated generation and evaluation errors if both stages rely on similar model assumptions.

A further limitation is that the present evaluation does not compare enrichment against established schema-matching baselines such as lexical/structural matching, graph-based similarity, composite matchers, or modern retrieval-plus-reranking approaches [20]–[27]. The reported LLM screening rates therefore establish the behavior of the implemented enrichment stage but not its relative advantage over non-generative or hybrid alternatives. Future work should include such baselines and evaluate precision, recall, candidate-reduction rate, review effort, and calibration on an independently annotated subset.

The strongest supported conclusions are therefore operational and structural: the pipeline completed successfully on 244 of 249 evaluated inputs, produced inspectable intermediate artifacts, preserved table and column inventory through the evaluated transformations, distinguished declared, derived, and suggested elements through explicit provenance metadata, and exposed sources of uncertainty rather than silently materializing uncertain structure. The results do not establish semantic correctness of recovered primary keys, inferred relationships, or conceptual interpretations. Claims about semantic validity require an independent reference set, deterministic validation of candidate keys and relationships, expert adjudication of domain meaning, and evidence that agreement reflects the true schema rather than systematic propagation of parser limitations or heuristic assumptions.

To make this evidence boundary explicit, Table VII summarizes which reconstructed properties are established by the present experiment and which require additional independent validation.

Accordingly, table and column preservation are supported directly by the evaluated artifacts, whereas heuristic keys, model-suggested relationships, conceptual names, and business rules remain dependent on additional schema- or domain-level validation.

TABLE VII. EVIDENCE STATUS OF RECONSTRUCTED PROPERTIES AND REQUIRED VALIDATION.

| Property | Evidence source | Current evaluation | Status | Required validation |
|---|---|---|---|---|
| Table existence | DDL | Table preservation | Declared | — |
| Column existence | DDL | Column preservation | Declared | — |
| Physical type | DDL | Type agreement | Declared | Type mapping review |
| Primary key from DDL | DDL | PK agreement | Declared | Independent reference |
| PK from heuristic | Identity/ naming rules | Backfill coverage | Derived | Independent reference |
| Declared FK | DDL constraint | FK count / graph | Declared | Reference or consistency checks |
| FK candidate | LLM | Parent-table admissibility | Suggested | Key/column/domain validation |
| Cardinality | Nullability/ uniqueness | Derived | Derived | Independent schema/domain evidence |
| Conceptual entity name | LLM | Not independently validated | Suggested | Expert/domain review |
| Business rule | LLM/ context | Not independently validated | Suggested | Domain review |

## V. LIMITATIONS

The corpus contains 249 heterogeneous Oracle-oriented samples and 1,225 SQL files, but its provenance, domain distribution, and sampling procedure are not fully documented. Thus, generalization to other Oracle databases, other DBMS platforms, production conventions, and less common DDL constructs remains unknown. Five samples failed (2.01%), chiefly because of parser-library edge cases involving ALTER TABLE, and the reported metrics describe only the 244 completed samples. The 52 completed samples for which the investigation stage produced no extractable CREATE TABLE or ALTER TABLE content and the possibility that failed samples are systematically more difficult further limit claims about robustness; future evaluations should compare completed and failed samples by observable complexity and report failure categories alongside quality metrics.

The post-backfill catalog used as the evaluation reference is an internal reconstruction, not an independently annotated semantic oracle. It inherits parser omissions, type-mapping assumptions, normalization choices, and heuristic primary-key recovery, while source DDL may omit application-level or conceptual constraints. Similarly, the reported 17.9% and 17.5% values are parent-table admissibility rates, not foreign-key accuracy or precision: they test only whether the suggested parent table exists; they do not verify column compatibility, parent-key membership, direction, cardinality, modality, or business meaning. Sample-level quality metrics are reported as macro-averaged point estimates, whereas table, key, and relationship counts are aggregate corpus totals over heterogeneous completed samples, without dispersion measures, confidence intervals, micro-averages, or paired significance tests. Consequently, structural agreement and candidate-generation rates should not be interpreted as semantic correctness. Independent expert-annotated references, deterministic candidate checks, and per-sample statistical analysis are required for stronger claims.

Parser coverage is bounded by supported Oracle syntax, identifier conventions, cross-file references, and indirect constraints, while naming heuristics can miss or falsely infer composite or domain-specific keys. LLM-dependent stages additionally require externally provisioned Azure OpenAI credentials and may vary with service updates, prompt changes, infrastructure, or model drift even at temperature zero; the model-assisted evaluator also creates a risk of correlated errors. The pipeline does not systematically redact sensitive DDL, document service retention, or provide a complete compliance audit trail, and locally hosted or hybrid alternatives were not evaluated. Reported LLM results should therefore be tied to the recorded model, API version, temperature, and execution context, and treated as reviewable operational evidence rather than verified schema semantics.

## VI. Conclusions

This study presented and evaluated a deterministic-first pipeline for reconstructing logical and conceptual data specifications from Oracle-oriented DDL. The pipeline separates investigation, parsing, consolidation, deterministic completion, declared-graph construction, and optional LLM enrichment. Its central representational decision is to preserve declared and deterministically derived structure in the baseline catalog and graph, while recording model-generated primary-key and foreign-key candidates in a separate overlay. This separation makes missingness and uncertainty explicit and prevents hypotheses from being silently promoted to schema facts.

The artifactized evaluation demonstrates that the approach is operationally viable for the examined corpus. The pipeline completed 244 of 249 samples (97.99%), while 52 completed samples contained no extractable table DDL and 192 produced non-empty reconstructed schemas. Across completed samples, the deterministic artifacts contained 208 reconstructed tables and 278 declared foreign-key records. These results support the use of the pipeline as an auditable structural baseline: intermediate catalogs, graph edges, source-file links, issues, and stage-specific error artifacts make failures and information loss inspectable. They do not, however, establish that the recovered catalog is a complete or semantically authoritative representation of the original systems.

The enrichment results reinforce the need for this conservative architecture. After deterministic backfill, 163 of the 192 completed samples with a non-empty reconstructed schema (84.90%) still contained at least one unresolved primary-key gap, showing that unresolved primary-key information remains substantial after deterministic reconstruction. LLM enrichment generated 100 foreign-key candidates in 36 samples, but only 17.9% of logical candidates and 17.5% of conceptual candidates named a parent table present in the reconstructed reference catalog. Because parent-table presence is only a necessary preliminary condition, these values are therefore parent-table admissibility measures rather than estimates of semantic foreign-key precision. They instead show that model-generated candidates that pass only limited preliminary screening are unsuitable for automatic materialization and should be subjected to deterministic checks and human or domain-expert review.

With respect to the central research objective, the results show that, for the successfully reconstructed non-empty cases in the evaluated corpus, Oracle-oriented DDL can be transformed into structurally preserving, provenance-aware logical and conceptual specifications, but semantic verification requires an independent validation source. The pipeline provides a provenance-preserving intermediate representation, combines deterministic recovery with controlled model assistance, and exposes declared, derived, and suggested elements as distinct classes. It also provides measurable execution and structural coverage over a heterogeneous corpus.

The evaluation does not answer the separate question of semantic correctness: the post-backfill catalog is an internal reconstruction reference rather than an independently annotated ground truth, and the evaluation lacks candidate-level validation of key membership, column compatibility, relationship direction, cardinality, and business meaning.

Accordingly, the principal contribution is a provenance-preserving reconstruction architecture that converts potentially incomplete physical schema evidence into reviewable logical and conceptual specifications without conflating deterministic evidence with generative hypotheses. In practical migration workflows, the deterministic catalog and declared graph can therefore be used as the migration baseline for schema inventory, impact analysis, and gap localization, whereas the enrichment overlay should be used only as a review-support mechanism for prioritizing unresolved PK/FK hypotheses.

Future work should add independently reviewed reference schemas, deterministic validation of suggested relationships, per-sample uncertainty and dispersion statistics, targeted tests for unsupported Oracle syntax, and comparisons with alternative models and deployment modes. Until independently validated, generated conceptual interpretations and model-suggested keys or relationships should remain explicitly labeled as candidates and should not be promoted to source-grounded migration requirements without deterministic or expert validation.

## Declaration on Generative AI

During the preparation of this work, the authors used ChatGPT (OpenAI) to check grammar and spelling. After using this tool, the authors reviewed and edited the output as needed and take full responsibility for the content of the publication.